\documentclass[12pt,twoside,a4papper]{article}
\usepackage{graphics}
\usepackage{graphicx}
\usepackage{amssymb}

\begin{document}

\title{The Pioneer anomaly as acceleration of the clocks}
\author{Antonio F. Ra\~nada\\Facultad de F\'{\i}sica, Universidad Complutense,\\
E-28040 Madrid, Spain\thanks{E-mail afr@fis.ucm.es}\\ (To be
published in {\it Foundations of Physics})}
\date{1 September 2004}
\maketitle

\begin{abstract}
This work proposes an explanation of the Pioneer anomaly, the
unmodelled and as yet unexplained  blueshift detected in the
microwave signal of the Pioneer 10 and other spaceships by
Anderson {\it et al} in 1998. What they observed is similar to the
effect that would have either (i) an anomalous acceleration
$a_{\rm P}$ of the ship towards the Sun or (ii) an acceleration of
the clocks $a_{\rm t}=a_{\rm P}/c$. The second alternative is
investigated here, with a phenomenological model in which the
anomaly is an effect of the background gravitational potential
$\Psi (t)$ that pervades all the universe and is increasing
because of the expansion. It is shown that $2a_{\rm t}={\rm d}\Psi
/{\rm d}t ={\rm d}^2\tau _{\rm clocks} /{\rm d}t^2$, evaluated at
present time $t_0$, where $t$ and $\tau _{\rm clocks}$ are the
coordinate time and the time measured by the atomic clocks,
respectively.  The result of a simple estimate gives the value
$a_{\rm t}\simeq 1.8\times 10^{-18}\mbox{ s}^{-1}$, while Anderson
{\it et al} suggested $a_{\rm t}= (2.9\pm 0.4)\times
10^{-18}\mbox{ s}^{-1}$ on the basis of their observations. The
calculation are performed near the Newtonian limit but in the
frame of general relativity.
\end{abstract}


\section{INTRODUCTION}  \paragraph{1.1 The anomaly.}
Anderson {\it et al} reported in 1998 the observation of an
unmodelled Doppler blueshift in the microwave signals from the
Pioneer 10/11, Galileo and Ulysses spacecrafts that increases
linearly in time$^{(1)}$. They had been observing it since more
than twenty years. Obviously, its simplest interpretation is that
the ships were not following the predicted orbits, as if our star
pulled a bit too much from them with a force independent of the
distance. The corresponding anomalous acceleration, directed
towards the Sun and constant, would have the value$^{(2)}$
\begin{equation}a_{\rm P}=
(8.74\pm 1.33)\times 10^{-10}\mbox{m/s}^2\,
.\label{1.1}\end{equation} Intriguingly enough, the effect does
not show up in the planets. In their first paper they said: ``it
is interesting to speculate on the possibility that the origin of
the anomalous signal is new physics"$^{(1,\,2)}$, and later
 ``The veracity of the signal is now undisputed, although
the source of the anomaly, some systematic or some not understood
physics, is subject to debate"$^{(3)}$. For an interesting
argument showing that it may not necessarily be due to
systematics, see reference 4. The effect is still
unexplained$^{(2)}$.

  Anderson {\it et al} say that their data show$^{(1)}$ ``a steady
frequency drift of about $-6\times 10^{-9}\mbox{ Hz/s}$, or 1.5 Hz
over 8 yr. This equates a clock acceleration, $-a_{\rm t}$, of
$-2.8\times 10^{-18}\mbox{ s/s}^2$", what would mean that the
frequencies would drift as \begin{equation} \nu =\nu _0[1+2a_{\rm
t}(t-t_0)]\,,\label{1.2}\end{equation} $t_0$ being here the
initial time of the observations (with the best value for $a_{\rm
P}$ (\ref{1.1}), $a_{\rm t}=(2.9\pm 0.4)\times 10^{-18}\mbox{
s}^{-1}$).
 The relation with the Pioneer
acceleration is $a_P\equiv a_tc$. This is important since they say
that the drift in the Doppler residuals cannot be removed without
either an acceleration of the ship $a_P$ or the inclusion of a
``clock acceleration" $a_{\rm t}$. Such acceleration $a_{\rm t}$
would imply that all the clocks would be changing with a constant
acceleration or, in other words, that there would be a
nonuniformity of time. They found that the first alternative leads
to problems with the equivalence principle and with the
cartography of the solar system. They considered the second by
means of several models in which the time is distorted
phenomenologically (see reference 2, sections XI.D and XI.E). The
best results were obtained with a model that adds a quadratic term
to the definition of the International Atomic Time. However, they
found some problems and concluded: ``The orbit determination
process clearly prefers the constant acceleration model, $a_{\rm
P}$, over the quadratic in time model."

\paragraph{1.2 Purpose and assumptions of this work.}
This paper considers this second alternative. It shows that,
because of the expansion, the background gravitational potential
that pervades all the universe produces an acceleration of the
cosmological proper time with respect to the coordinate time. In
its turn, this implies an acceleration $a_{\rm t}$ of the atomic
clocks. A simple estimation gives a value for $a_{\rm t}$ that is
close to that found by Anderson {\it et al}. The anomaly is thus
in this model an effect of the expansion of the universe. It is
assumed, for simplicity, that (i) all the matter and energy of the
universe are uniformly distributed, (ii) the space sections $t=$
constant are flat, and (iii) the near Newtonian approximation is
adequate and meaningful. The time coordinate $t$ is so chosen as
to go in the Newtonian limit to the Newtonian time. The model here
presented is the relativistic version of a previous Newtonian
one$^{(5-7)}$, see also reference 8. Two other models in which the
anomaly is also due to the expansion are proposed in references 9
and 10.

\paragraph{1.3 Two definitions of the light speed.}
It is important to know precisely which one of the several
meanings of ``speed of light" is used$^{(11)}$. Particularly, it
must be reminded here that the light speed can be defined in
general relativity in two different ways, (i) with respect to the
cosmological proper time $\tau$, $c^*={\rm d}\ell /{\rm d} \tau \,
(=\mbox{constant})$, and (ii) with respect to the coordinate time
$t$, $c ={\rm d}\ell /{\rm d} t=c({\bf r}, t)$, where ${\rm
d}\ell$, ${\rm d} \tau$ and ${\rm d} t$ are elements of spatial
distance, proper time and coordinate time along a null geodesic,
respectively (the first is the usual definition). These two speeds
will be denoted as $c^*$ and $c$, respectively, and will be called
``{\it proper speed of light}" and ``{\it ordinary} or {\it
non-proper speed of light}. The derivative of $c$ with respect to
$t$ at present time $t_0$, denoted as $a_\ell=\dot{c}(t_0)={\rm
d}c(t_0)/{\rm d}t$, will be called {\it non-proper acceleration of
light} or just acceleration of light if there is no risk of
confusion. The first is a universal constant of nature (as it must
happen in general relativity), the second is not but, quite on the
contrary, it depends generally on space and time $c=c({\bf r},t)$.
The duality between $c^*$ and $c$ reflects the relation between
the proper time and the coordinate time.

The element of interval can be written$^{(12-15)}$ (assuming for
simplicity that $g_{0i}=0$)
\begin{equation}
{\rm d}s^2=c^{*\, 2}{\rm d}\tau ^2-{\rm d}\ell ^2\,
,\label{1.3}\end{equation} with ${\rm d}\tau =\sqrt{g_{00}}\,{\rm
d}t$ and ${\rm d}\ell ^2=g_{ij}{\rm d}x^i{\rm d}x^j$, so that
$c^*$ is constant  and $c$ is equal to
\begin{equation}
c=c({\bf r},t)=c^* \sqrt{g_{00}}\, .\label{1.4}\end{equation} Near
the Newtonian limit, $g_{00}\simeq 1+2\Phi /c^2$, at first order,
$\Phi$ being the gravitational potential, so that $c=c({\bf
r},t)=c^*[1+\Phi ({\bf r},t)/c^2].$ Taking a non-zero origin for
the potential at a reference laboratory $R$, this can be written,
at first order, as
\begin{equation}
c=c({\bf r},t)=c_0[1+\Phi ({\bf r},t)/c^2({\bf r},t)-\Phi
_R/c_0^2]\, ,\label{1.6}\end{equation} where $c_0$ and $\Phi _R$
are the values of $c$ and the potential at that laboratory.

In this paper, we are interested in the effect of the background
gravitational potential that pervades  all the universe and is due
to all the existing matter and energy. Assuming a uniform
distribution of matter and energy, it is clear that  is time
depending but space independent. It will be denoted as $\Phi _{\rm
av}(t)$ (``av" stands for average since the mass-energy density is
averaged). The same must happen therefore to $c=c(t)$. Instead of
(\ref{1.6}) one would have then
\begin{equation}
c=c(t)=c_0[1+\Phi _{\rm av}(t)/c^2(t)-\Phi _{\rm av}(t_0)/c_0^2]\,
,\label{1.6b}\end{equation} with $c_0=c(t_0)$, $t_0$ being a
reference time that will be the present time or age of the
universe, in general. Because of the expansion, it turns out that
$\Phi _{\rm av}(t)/c^2(t)$ is an increasing function, as will be
shown in section 4 where its derivative with respect to $t$ will
be calculated. Consequently, $c(t)$ is also increasing.

If the universe would contain only matter, be it ordinary or dark,
the background potential $\Phi _{\rm av}$ would be negative so
that
\begin{equation}
c(t)< c_0[1-\Phi _{\rm av}(t_0)/c_0^2]\,
.\label{1.bc}\end{equation} As will be seen in section 3, this
implies $c(t)< c^*$, as it could be expected.

However, the effect of the cosmological constant or of dark energy
changes dramatically this question. The cosmological model used
here is the standard with $27\,\%$ of matter and $73\,\%$ of dark
energy. For the latter we take either a cosmological constant or
the quantum vacuum, but the conclusion would be the same for any
kind of dark energy with an  equation of state  implying
repulsion. The potential $\Phi _{\rm av}(t)$ is the addition of
the two effects of matter and dark energy.

What is important here is that both the cosmological constant and
the quantum vacuum produce a positive potential. Furthermore, it
turns out that $\Phi _{\rm av}(t)$ is an increasing positive
function after a certain time, in particular now (this will be
shown in section 4). Equation (\ref{1.6b}) implies then that $c$
increases  as far as the universe expands and, in particular, it
can be larger than the proper speed of light $c^*$. Although this
maybe seem strange and contrary to current wisdom, it must not be
a matter of concern since, as it must be emphasized, the proper
light speed of light $c^*$ is in fact a universal constant in this
model. Moreover, $c< c^*$ if there is only matter, let it be
ordinary or dark. It is only the dark energy, which, in addition
to accelerate the universe, can make that $c(t)$ could be larger
than the proper speed of light $c^*$. This intriguing result will
be considered in section 3. To understand this property, one must
keep in mind that almost all our intuitions in general relativity
were generated in the study of the gravitation of matter ({\it i.
e.} without dark energy).

There is moreover a functional relation between the two times
$\tau =\tau (t)$, such that $\tau$ must accelerate with respect to
$t$. Because of the freedom to choose the coordinates in general
relativity, this last statement must be qualified: the time
coordinate $t$ is defined here so that in the Newtonian limit it
goes over the Newtonian time. All this means that
\begin{equation}
c(t)={{\rm d}\ell \over {\rm d}\tau}\,{{\rm d}\tau \over {\rm
d}t}= c^*\,{{\rm d}\tau \over {\rm d}t} \qquad \mbox{and}\qquad
{{\rm d}c(t)\over {\rm d}t}=c^*{{\rm d}^2\tau \over {\rm d}t^2}\,
.\label{1.8}\end{equation}  Equation (\ref{1.8}) states that {\it
the non-proper light speed $c(t)$ must increase if the proper time
$\tau$ accelerates with respect to the coordinate time $t$}, its
time derivative being equal to the proper light speed $c^*$ times
the second derivative of $\tau$ with respect to $t$.

The expression ``light speed" means usually the proper speed $c^*$
that, being a universal constant, is of the utmost importance.
However, the non-proper light speed $c$ is also used in some
important cases. For instance, in the study of the bending of a
light ray grazing the Sun surface.  Let $M$ and $R$ be the mass
and radius of the Sun. The interval around any star is given by
the Schwarzschild metric, what implies that $c= c(r)= c_\infty
(1-\eta R/r)$, with $c_\infty =c(\infty)$ and $\eta
=GM/c_\infty^2R\simeq 2.1 \times 10^{-6}$. Einstein gave two
formulae for this effect. The first (1907) is based only in the
equivalence principle and gives $\phi =2\eta = 0.875^{\prime
\prime}$, just one half of the observed effect. The second (1916),
in the frame of general relativity, gives the complete result
$\phi =4\eta = 1.75^{\prime \prime}$. The first one can be
obtained simply by considering the propagation of a wave light
with the previous value of the non-proper light speed, in other
words as the solution of the variational problem
\begin{equation}
\delta T=\delta \int _1^2{1+\eta R/r\over c_\infty}\, {\rm d}\ell
 =0\, ,\label{1.3.5}\end{equation} where ${\rm d}\ell={\rm
d}x^2+{\rm d}y^2+{\rm d}z^2$ is the Euclidean line element, {\it
i.e.} as a consequence of the application of the Fermat principle
to the non-proper light speed $c$. The complete effect is obtained
by taking instead the non-Euclidean spatial line element of the
Schwarzschild geometry. In other words, the problem is solved by
assuming that the light propagates through space with the
non-proper speed $c(r)= c_\infty (1-\eta R/r)$ (taking into
account the Riemannian character of the spatial metric).

  These
considerations can be summarized as follows: because of the
background gravitational potential of all the matter and energy in
the universe, (i) the cosmological proper time $\tau$ accelerates
with respect to the coordinate time, and (ii) the non-proper speed
of light $c(t)$ increases so that $a_\ell>0$.

\paragraph{1.4 Plan of the paper.} In section 2, the Maxwell equations
in general relativity are reviewed, with emphasis on the ideas of
permittivity and permeability of geometrical origin and on their
effect on the non-proper speed of light $c({\bf r},t)$. It is
shown that $2a_{\rm t}$ must be equal to the derivative with
respect to $t$ of the background potential over $c^2(t)$ and that
the corresponding non-proper of light acceleration $a_\ell$
implies a blueshift. In section 3, the reasons for the
acceleration of the atomic clocks are considered, showing that it
is equal to $2a_{\rm t}$. In section 4, an estimation of $a_{\rm
t}$ is carried out, the agreement being good with the result found
by Anderson {\it et al} on the basis of their
observations$^{(1)}$.  The conclusions will be stated in section
5.

\section{THE SPEED OF LIGHT \\ AND THE MAXWELL EQUATIONS}
In order to understand the behavior of the non-proper speed of
light $c({\bf r},t)$ (different, don't forget, from the  proper
speed of light $c^*$, which is a universal constant), let us
consider the effect of a gravitational field on the Maxwell
equations, in which the time derivatives are with respect to the
coordinate time $t$. As was seen in section 1.3, the proper speed
of light $c^*$ and the non-proper speed of light $c$ are equal in
absence of potential or, in other words, the reason for their
difference is the presence of matter and energy. It is clear that
near the special relativity or the Newtonian limits, the
electromagnetic fields obey the classical wave equation with the
velocity $c({\bf r},t)$ (remember that flat space sections are
assumed). That local value of $c$ must be used at any spacetime
point in the wave equation. Let us take a reference laboratory
where $c=c_0$. The following discussion is based on the well known
textbook {\it The Classical Theory of Fields} by Landau and
Lifshitz (reference 16, section 90).

The electromagnetic tensor is defined in general relativity by
means of a vector field such that $ F_{\mu
\nu}=A_{\nu;\mu}-A_{\mu;\nu}=\partial _\mu A_\nu -\partial _\nu
A_\mu \, .$  The electromagnetic vectors $\bf E$, $\bf D$ and
antisymmetric tensors $B_{ij},\, H_{ij}$ are defined as follows
$$E_i=F_{0i}\, ,\quad B_{ij}=F_{ij}\, ,\quad
D^i=-\sqrt{g_{00}}\, F^{0i}\, ,\quad H^{ij}=\sqrt{g_{00}}\,
F^{ij}\, ,$$ the vectors $\bf B$, $\bf H$ being the dual to the
three-tensors $B_{ij}$ and $H_{ij}$, {\it i. e.} $B^i=-
e^{ijk}B_{jk}/(2\sqrt{\gamma})$, $H_i=-\sqrt{\gamma}\,
e_{ijk}H^{jk}/2$, where $\gamma =\det (\gamma _{ij})$, $\gamma
_{ij}= -g_{ij}$ being the three-dimensional metric tensor
(assuming for simplicity $g_{0i}=0$). It follows that$^{(16)}$
($\epsilon _0=1$, $\mu _0=1$ in this argument)
\begin{equation} {\bf D}={\bf E}/ \sqrt{g_{00}}\, , \quad {\bf
B}={\bf H}/ \sqrt{g_{00}}\, .\label{2.1}\end{equation}  If the
space is empty, {\it i. e.} without free charges or currents, the
Maxwell equations can be written as
\begin{equation} \nabla \cdot {\bf B}=0\, , \quad \nabla \times
{\bf E}=-{1\over \sqrt{\gamma}}\, \partial _t\left(
\sqrt{\gamma}\, {\bf B}\right)\, .\label{2.2}\end{equation}
\begin{equation}
\nabla \cdot {\bf D}=0\, ,\quad \nabla \times {\bf H}= {1\over
\sqrt{\gamma}}\,
\partial _t\,\left(\sqrt{\gamma}\,{\bf D}\right)\, .\label{2.3}\end{equation} In a static
situation, these four equations have exactly the same form as in
special relativity, since the factors $\sqrt{\gamma}$ cancel.
However, eq. (\ref{2.1}) implies that the relative permittivity
$\epsilon _{\rm r}$ and permeability $\mu _{\rm r}$ of empty space
are different from 1, their common value being $\epsilon _{\rm
r}=\mu _{\rm r}=(g_{00})^{-1/2}$. This is due to the geometry of
spacetime. If the potential depends on time and near the Newtonian
limit one has $g_{00}=1+\Phi ({\bf r},t)/c^2({\bf r},t)-\Phi _{\rm
R}/c_0^2$, where $\Phi _{\rm R}$ is the potential at a reference
laboratory where $c=c_0$, so that the empty space is like an
inhomogeneous optical medium with
\begin{equation}\epsilon _{\rm r}({\bf r},t)=\mu _{\rm r} ({\bf r},t) =1-[\Phi
({\bf r},t)/c^2({\bf r},t)-\Phi _{\rm R}/c_0^2]\,
.\label{2.4}\end{equation} Since $c=c_0/\sqrt{\epsilon _{\rm r}\mu
_{\rm r}}$, the non-proper speed of light is given as
\begin{equation}
c({\bf r},t ) =c_0\left\{1+\Phi ({\bf r},t)/c^2({\bf r},t)-\Phi
_{\rm R}/ c_0^2\right\}\, , \label{2.5}
\end{equation}
$\Phi _{\rm R}$ being a reference potential, at present time in a
terrestrial laboratory $R$, where the observed light speed is
$c_0$. As a historical comment, this last equation was first
obtained in 1911 by Einstein himself, as a first order
approximation in the static case, in a paper entitled ``On the
influence of gravitation on the propagation of light"$^{(17)}$ (in
1907 he had already shown that $c$ depends on $\Phi$ as a
consequence of the equivalence principle. Note that it is still
valid in general relativity, at first order.) After a discussion
on the synchronization of clocks, he concludes there ``if we call
the velocity of light at the origin of coordinates $c_0$, where we
take $\Phi =0$, then the velocity of light at a place with
gravitational potential $\Phi$ will be given as
\begin{equation}
c=c_0\left(1+\Phi / c^2\right)\, ." \label{2.6} \end{equation}
Einstein had not as yet introduced  the general relativistic idea
of proper time and used only the coordinate time to define the
speed of light$^{(18-19)}$. Note that (\ref{2.5}) reduces to
(\ref{2.6}) in the static case if the potential at the reference
laboratory vanishes and that the consideration of the Maxwell
equations confirms eq. (\ref{1.6}), obtained in section 1.3 from
the expression of the interval.

In this work, we are interested in the case of a potential
depending only on time. Let $\Phi _{\rm av}(t)$ be the background
potential of all the matter and energy and $\Psi (t)=\Phi _{\rm
av}(t)/c^2(t)$ its dimensionless expression (the subindex ``av"
being omitted in $\Psi$ in order to simplify the notation).
 The permittivity and permeability (\ref{2.4}) take then the
 form near present time $t_0$ ({\it
i.e.} the age of the universe)
 \begin{equation}\epsilon _{\rm r}(t)=\mu _{\rm r}(t) =1-[\Psi
 (t)-\Psi (t_0)]\, .\label{2.7}\end{equation}
 The non-proper speed of light is then
\begin{equation}
c(t)=c_0\left[1+\Psi(t)-\Psi (t_0)\right]\, , \label{2.8}
\end{equation}
where $c_0=c(t_0)$. This can be written as
\begin{equation} c(t)=c_0[1+2a_{\rm t} (t-t_0)]=c_0+a_{\rm
\ell}(t-t_0), \label{2.9}\end{equation} the quantity $a_{\rm t}$
and the non-proper acceleration of light $a_\ell=\dot{c}(t_0)$
being
\begin{equation}   2a_{\rm t} = \dot{\Psi} (t_0)\, ,\quad \quad
 a_\ell =2a_{\rm t}c_0=2\dot{\Psi}(t_0)c_0\, .\label{2.10}
\end{equation}
As will be seen in the following, (i) $a_\ell$ is in fact $2a_{\rm
P}$, being therefore an adiabatic acceleration, and (ii) $a_t$ is
what Anderson {\it et al} termed acceleration of the clocks.

\paragraph{2.1 The blueshift.} It will be shown now that the non-proper acceleration
$a_\ell$ implies a blue shift of the light with respect to the
coordinate time, at first order in $a_\ell$. More precisely, it
turns out that the frequency $\nu $ of a monochromatic light wave
with such an adiabatic acceleration $a_\ell$ increases, its
derivative with respect the coordinate time $t$, $\dot{\nu}$,
satisfying
\begin{equation}
\dot{\nu}/ \nu_0 =a_\ell/c_0. \label{2.11}
\end{equation}
This means that an adiabatic non-proper acceleration of light has
the same radio signature as a blue shift of the emitter, although
a peculiar blue shift with no change of the wavelength ({\em i.e.}
all the increase in velocity is used to increase the frequency).

 The derivative with respect to $t$ of the background gravitational
 potential of all the universe  $\Psi (t)$ is
 positive and of the order of the Hubble parameter
 $H_0=2.3\times 10^{-18}\mbox{ s}^{-1}$, since the galaxies are separating (a calculation will be
 made in section 4).
 Equation (\ref{2.7}) tells then that the relative permittivity
 $\epsilon _{\rm r}$ and permeability
 $\mu _{\rm r}$ of empty space are decreasing,
  their derivatives with respect to $t$
 being negative and also of order $H_0$, {\it i. e.} very small.
 This can be expressed by saying that the optical
 density of empty space is decreasing adiabatically. To study the
propagation of the light in a medium with time depending
permittivity and permeability, we must take the Maxwell equations
and deduce the wave equations for the electric field ${\bf E}$ and
the magnetic intensity $\bf H$, which are $\nabla ^2{\bf
E}-{\partial _t}\left( \mu {\partial _t}(\epsilon {\bf
E})\right)=0$, $\nabla ^2{\bf H}-{\partial _t}\left( \epsilon
{\partial _t}(\mu {\bf H})\right)=0,$ or, more explicitly,
\begin{eqnarray}
\nabla ^2{\bf E}-{\partial _t^2}{\bf E}/c^{2}(t) &-&
\left({\dot{\mu}/ \mu _0}+{2\dot{\epsilon}/\epsilon
_0}\right){\partial _t{\bf E}}/c^2(t)
-{\dot{\epsilon}\dot{\mu}}{\bf E}/(\epsilon _0\mu _0c^2(t))=0,\nonumber \\ &&\label{2.12}\\
\nabla ^2{\bf H}-{\partial _t^2{\bf H}}/ c^2(t) &-&
\left({2\dot{\mu}/\mu _0}+{\dot{\epsilon}/\epsilon
_0}\right){\partial _t{\bf H}/ c^2(t)}
-{\dot{\epsilon}\dot{\mu}{\bf H}/( \epsilon _0\mu _0 c^2(t))}
=0,\nonumber \\ \label{2.13}
\end{eqnarray}
with $c(t)=c_0+a_{\rm t}(t-t_0)$, since at present time $\epsilon
_{\rm r}=1,\;\mu _{\rm r}=1$. Because $\dot{\epsilon}/\epsilon _0$
and $\dot{\mu}/\mu _0$ are of order $H_0=2.3\times 10^{-18}\mbox{
s}^{-1}$, the third and the fourth terms in the LHS of
(\ref{2.12}) and (\ref{2.13}) can be neglected for frequencies
$\omega \gg H_0$, in other words for all practical purposes.  We
are left with two classical wave equations with time dependent
light velocity $c(t)$.
\begin{equation}
\nabla ^2{\bf E}-{\partial _t^2{\bf E}/ c^2(t)}=0,\;\; \nabla
^2{\bf H}-{\partial _t^2{\bf H}/ c^2(t)}=0. \label{2.14}
\end{equation}
In order to find the behavior of a monochromatic light beam
according to these two wave equations, we take for instance the
first one and insert ${\bf E}={\bf E}_0\exp\{{-i[\kappa z-(\omega
_0+\dot{\omega}(t-t_0)/2)(t-t_0)]}\}$, where the frequency is the
time derivative of the phase of {\bf E}, {\em i.e.} $\omega
_0+\dot{\omega}(t-t_0)$. Neglecting the second time derivatives
and working at first order in $\dot{\omega}$ (with
$\dot{\omega}(t-t_0)\ll \omega_0$, $\dot{\omega}\ll \omega _0^2$),
substitution in (\ref{2.12}) gives $\kappa ^2 =\left[(\omega_0
+\dot{\omega}(t-t_0))^2-i\dot{\omega}\right]/c^2(t)$. It follows
that $\kappa =k+i \zeta = \pm (\omega_0/
c(t))[1+\dot{\omega}(t-t_0)/\omega_0](\cos \varphi + i \sin\varphi
),$ with $\varphi =-\dot{\omega}/2\omega_0 ^2,$ so that $k=\pm
(\omega_0 /c_0)\, (1+\dot{\omega}(t-t_0)/ \omega_0)/(1+a_\ell
(t-t_0)/ c_0)$ what implies $k=\pm \omega _0/c_0$, $\dot{\omega}
/\omega_0= a_\ell/c_0$, as stated before. Equation (\ref{2.11})
has thus been proved. Also, $\zeta =-\dot{\omega}/2\omega_0 c_0
=a_\ell /2c_0^2$. The wave amplitude decreases in the direction of
propagation as $e^{-z/\ell}$ with $\ell=2c_0^2/a_\ell$, but as
$a_\ell $ is of order $H_0c_0$, $\ell$ is of order of 5,000 Mpc,
or, in other words, this attenuation can be neglected. As is easy
to show, to take $k +\dot{k}t$ for the wave vector leads to
$\dot{k}=0$. These results are equally valid for the second
equation in (\ref{2.14}). Taking into account that $a_\ell /c_0=
2a_{\rm t}$ and according to (\ref{2.7})-(\ref{2.10}), the
frequencies drift as
\begin{equation}
\nu =\nu _0[1+2a_{\rm t}(t-t_0)]\, ,\label{2.15}\end{equation}
what shows that $2a_t=\dot{\Psi }(t_0)$ is the acceleration of
clocks mentioned by Anderson {\it et al}. According to these
arguments, its value is the derivative with respect to the
coordinate time $t$ of the background gravitational potential of
all the universe. This will be further studied in next section.

\section{THE ACCELERATION OF THE ATOMIC CLOCKS}

\paragraph{3.1 The effect of the dark energy. Can $c(t)$ be larger
than $c^*$?} The observations are made by using atomic clocks,
which measure proper time $\tau$ not coordinate time $t$.
However, up to now we have used mainly in this paper the
coordinate time. This question is  addressed in this section.
Taking the $t$ derivative of (\ref{2.8}) at time $t_0$, it is
found that $\dot{c}(t_0)=c(t_0)\dot{\Psi}(t_0)$. The same argument
can be applied to the expression for $c(t)$ near any other fixed
time $\tilde{t}$, what implies that $c(t)=c(\tilde{t})\exp[\Psi
(t)-\Psi (\tilde{t})]$, an expression valid for $\forall t$. In
particular, taking $\tilde{t}=t_0$, one finds
\begin{equation}c(t)=c_0\,e^{[\Psi (t)-\Psi (t_0)]}\, .\label{3.2}\end{equation}
As can be seen, (\ref{2.8}) is the first order approximation to
({\ref{3.2}). The shape of the function $c(t)$ defined by
(\ref{3.2}) does not change if the reference time ($t_0$ or
$\tilde{t}$) is changed, because $c(t_1)e^{-\Psi
(t_1)}=c(t_2)e^{-\Psi (t_2)}$. Since $c^*$ and $c$ are equal if
$\Psi =0$, it follows that $c^*=c_0 e^{-\Psi (t_0)}$, what does
not depend on the particular value of $t_0$. In other words, the
non-proper speed of light is
\begin{equation}
c(t)=c^*e^{\Psi (t)}\, . \label{3.2b}\end{equation} We can precise
now some ideas exposed in section 1.3. If there is only matter in
the universe ({\it i. e.} without dark energy), then $\Psi (t)$ is
necessarily negative, so that $c(t)< c^*$. This is the usual
situation that everybody has in mind. However, if there is dark
energy, $\Psi (t)$ can be positive, and in that case $c(t)>c^*$.
We find thus that the non-proper speed of light could be larger
than the proper speed and constant of the nature $c^*$. Even if
this is a surprising effect of the dark matter, in addition to
accelerate the expansion, it must be emphasized that it is
compatible with the constancy of $c^*$ and with general relativity
therefore. Indeed, it is an unexpected and new effect of the dark
energy.

\paragraph{3.2 On the time of the atomic clocks.}

The interval can be written then as
\begin{equation}
{\rm d}s^2=e^{2[\Psi (t)-\Psi (t_0)]}c_0^2{\rm d}t^2-{\rm d}\ell
^2=c^{*\, 2}{\rm d}\tau ^2-{\rm d}\ell ^2\,
\label{3.3}\end{equation} so that\begin{equation} {\rm d}\tau
=e^{\Psi (t)}{\rm d}t \quad \mbox{ and } \quad \left.{{\rm
d}^2\tau \over {\rm d}t^2}\right|_{t_0}=\dot{\Psi} (t_0)e^{\Psi
(t_0)}=2a_{\rm t}e^{\Psi (t_0)}>0\, ,
\label{3.4}\end{equation}since $\Psi (t)$ is now an increasing
function and $\dot{\Psi}(t_0)>0$ (the precise calculation is done
in next section). Note that $\tau$ is a well defined cosmological
proper time and that it accelerates with respect to $t$, its
second derivative being in general non nil. This means that if,
sometime $t_{\rm i}$ in the past, an atomic clock was set to click
at the same rate as the coordinate time, it would be advanced now
with respect to the coordinate time. In particular, the two time
intervals would be
 different now, since ${\rm d}\tau =e^{\Psi (t_0)}{\rm d}t$. If
 there is only matter, be it ordinary or dark, then $\Psi (t_0)<0$
 and ${\rm d}\tau <{\rm d}t$. On the other hand, if the dark energy exists,
 then $\Psi (t_0)$ can be positive and then ${\rm d}\tau >{\rm d}t$.
 This second possibility is the one that actually happens: the background potential not only increases, but is
positive also because of the effect of the quantum vacuum, or the
dark energy, as is shown in next section.

However, in the actual measurements these differences between
${\rm d}t$ and ${\rm d}\tau$ do not occur, since both times are
based now on the international second, defined with reference to
the period of a transition of the cesium atom.   This means that a
small interval of the atomic clock time and of the coordinate time
are equal, ${\rm d}\tau _{\rm clocks}={\rm d}t$, at precisely
$t=t_0$. Therefore, the time of the atomic clocks has been
renormalized, indeed, multiplying the cosmological proper time by
the constant scale factor $e^{-\Psi (t_0)}$. This argument shows
that the time really measured by the atomic clocks is
\begin{equation} {\rm d}\tau _{\rm clocks}
=e^{[\Psi (t)-\Psi (t_0)]}{\rm d}t \quad \mbox{ so that } \quad
\left.{{\rm d}^2\tau _{\rm clocks} \over {\rm
d}t^2}\right|_{t_0}=\dot{\Psi} (t_0)=2a_{\rm t}\, .
\label{3.5}\end{equation} This explains the real meaning of
$a_{\rm t}$: {\it it is one half the second derivative with
respect to the coordinate time $t$ of the time of the atomic
clocks, if they have been renormalized by a multiplicative factor
to tick now, just at time $t_0$, at the same rate as the
coordinate time} (in order for both to use the same definition of
second). Note that there is a difference between the meanings of
the ``acceleration of the clocks": here it is $2a_{\rm t}$ while
for Anderson {\it et al.} it is $a_{\rm t}$. Of course, in the
future the atomic clocks will advance over $t$, {\it i. e.} ${\rm
d}\tau _{\rm clocks}>{\rm d}t$ if $t>t_0$.

 Taking the first order approximations near $t_0$, the intervals of both
times are related as ${\rm d}\tau _{\rm clocks}=[1+\Psi (t)-\Psi
(t_0)]{\rm d}t$. This is equal to ${\rm d}\tau _{\rm clocks}
=[1+\dot{\Psi} (t_0)(t-t_0)]{\rm d}t=[1+2a_{\rm t}(t-t_0)]{\rm
d}t$ (compare with (\ref{1.2})). It follows
\begin{equation}
\tau _{\rm clocks}-\tau _{\rm clocks} (t_0) =(t-t_0)+a_{\rm
t}(t-t_0)^2\, ,\label{3.6}\end{equation} what reminds the
quadratic in time model tried by Anderson {\it et al} to solve the
anomaly. Probably, this explains why this was the best among the
other models they used with phenomenologically distorted time (see
section 1.1, at the end).

 In order to refer to the time of the clocks, $a_{\rm
t}$ must be multiplied by the factor ${\rm d}t/{\rm d}\tau_{\rm
clocks}$ what produces the change
\begin{equation}
a_{\rm t}\Rightarrow a_\tau = a_{\rm t}{{\rm d}t\over {\rm d}\tau
_{\rm clocks}}=a_{\rm t}[1-2a_{\rm t}(t-t_0)]= a_{\rm t}\,
,\label{3.7}\end{equation} neglecting the terms of second order in
$a_{\rm t}$. Since $a_{\rm t}=a_{\tau}$, at first order, eq. (2)
can be written  as
\begin{equation} \nu =\nu _0\{1+2a_{\tau}[\tau _{\rm clocks}-
\tau_{\rm clocks}(t_0)]\}\,,\label{3.8}\end{equation}
at first order, what states that the frequency must drift,
according to the measurements made with atomic clocks. This gives
a solution to the riddle.

Indeed the two derivatives of the background potential $\Psi $
with respect to the times $t$ and $\tau _{\rm clocks}$ are equal
because the intervals of the two times are equal at precisely the
time $t_0$ (or $\tau _{\rm clocks}(t_0)$). Consequently, the
acceleration of the clocks can be calculated as one half  the
derivative of $\Psi$ with respect to any one of the two times (at
first order). Therefore, this model seems to offer an attractive
solution to the anomaly. But in order to test it, its prediction
for the inverse time $a_{\rm t}$ must be calculated and compared
with the value proposed by Anderson {\it et al}. This is done in
next section.

\section{ESTIMATION OF THE ACCELERATION OF THE CLOCKS}
 Unfortunately, a rigorous calculation of $a_{\rm t}$ that
would take into account all the eventual effects, is not easy.
However, a simple crude estimate of its value can indeed be made,
as is done in this section, taking (\ref{2.8}) and (\ref{2.10}) as
starting point. Although involving approximations and
simplifications, the result is meaningful since it shows the main
ideas of the model, giving a convincing representation of the
phenomenon. In order to do that, the shape of the function $\Psi
(t)$ near $t_0$ must be determined first to calculate its time
derivative eq. (\ref{2.10}). The potential of all the universe at
the terrestrial laboratory $R$ can be written, with good
approximation, as $\Phi _{\rm all}=\Phi _{\rm loc}(R)+\Phi_{\rm
av}(t)$. The first term $\Phi _{\rm loc}(R)$ is the part due to
the  local inhomogeneities, {\it i. e.} the nearby bodies (the
Solar System and the Milky Way). It is constant in time since
these objects are not expanding. The second $\Phi_{\rm av}(t)$ is
the space averaged potential due to all the mass and energy in the
universe (except for the nearby bodies), assuming that they are
uniformly distributed. Contrary to the first, it depends on time
because of the expansion. The former has a non vanishing gradient
but is small, the latter is space independent,
 but time dependent and much larger. The
 value of $\Phi _{\rm
loc}/c_0^2$ at the laboratory $R$ is the sum of the effects of the
Earth, the Sun and the Milky Way, which are about $-7\times
10^{-10}$, $-10^{-8}$ and $-6\times 10^{-7}$, respectively,
certainly with much smaller absolute values than $\Phi _{\rm
av}(t_0)$, of the order of $-10^{-1}$ as will be seen below.

Since the background gravitational potential of all the universe
$\Phi _{\rm av} (t)$ is increasing because of the expansion (the
galaxies are separating and their interaction potential
increasing), eqs. (\ref{2.10}) imply that $a_{\rm t}$ and $a_\ell$
are both positive, as will be seen in the following. In this
sense, there is a non-proper acceleration of light $a_\ell
=\dot{c}$, see section 1.3 (as it may be convenient to stress
again, all this is compatible with the constancy of the proper
speed of light $c^*={\rm d}\ell/{\rm d}\tau$.)

 Let $\Omega _{\rm M}=0.27$, $\Omega _\Lambda =0.73$ be the corresponding
present time relative densities of matter (ordinary plus dark) and
dark energy corresponding to the cosmological constant $\Lambda$.
We take a universe with $k=0$  and Hubble parameter $H_0=71\mbox{
km}\cdot \mbox{s}^{-1}\cdot \mbox{Mpc}^{-1}= 2.3\times
10^{-18}\mbox{ s}^{-1}$.
 In
order to determine the average potential $\Phi _{\rm av}(t)$, let
$\Phi _0(t_0)$ be the gravitational potential produced by the
critical density of mass distributed up to the present radius of
the visible universe $R_{\rm U}(t_0)=c_0/H_0= 4,200\mbox{ Mpc}$.
One has then $\Phi _0(t_0)/c_0^2=-\int _0^{c_0/H_0} c_0^{-2}G\rho
_{\rm cr}4\pi r{\rm d}r=2\pi G\rho_{\rm cr}/H_0^2\simeq -0.78$. It
must be emphasized that, although this value of the  potential
might seem to be too large for this approximation to apply, there
is no problem in fact since it is space independent and its time
derivative is extremely small. It will be absorbed in the
redefinition of time, its effect being only to accelerate
adiabatically the proper time with respect to the coordinate time,
{\it i.e.} to accelerate the clocks, precisely what we want to
investigate.

Because the radius of the universe is changing, the potential must
be multiplied by the factor $[R_{\rm U}(t)/R_{\rm U}(t_0)]^2$,
with $\dot{R}_{\rm U}(t_0)=c_0$. It turns out then that $\Phi
_0(t)/c_0^2=2\pi G\rho_{\rm cr}R_{\rm U}^2(t)/c_0^2$.

The present time value of the background potential is then $\Phi
_{\rm av}(t_0)=\Phi _0(t_0)(\Omega _M-2\Omega _\Lambda)\simeq
0.92>0$ (it is positive because of the contribution of the
cosmological constant). Because of the expansion of the universe,
the gravitational potentials due to matter and dark energy
equivalent to the cosmological constant vary in time as the
inverse of the scale factor $R(t)$ and as its square $R^2(t)$,
respectively (with $R(t)=\left({\Omega _M/ \Omega _\Lambda
}\right)^{1/3}\sinh ^{2/3}\left[{(3\Lambda )^{1/2}t/ 2}\right]$
for this model universe). This implies that the average background
gravitational potential can be expressed as
\begin{equation}
\Phi _{\rm av}(t)=\Phi _0(t_0)\left(R_{\rm U}(t)\over R_{\rm
U}(t_0)\right)^2\left[{\Omega _M\over R(t)}-2\Omega _\Lambda
R^2(t)\right]\, ,\label{4.1}\end{equation} (remember: $\Psi
(t)=\Phi _{\rm av}(t)/c^2(t)$). Note that, as announced in section
1.3 and since $\Phi _0(t_0)<0$, the term in $\Omega _\Lambda$
overcomes necessarily the one in $\Omega _{\rm M}$, after a
certain time, the potential being positive afterwards, as far as
the expansion goes on. The non-proper speed of light $c(t)$ is
then larger than $c^*$. As is easy to see, $\Phi _{\rm
av}(t)\rightarrow 0$ at time zero, because $R_{\rm U}(t)\sim t^2$
in that limit. After a bit of simple algebra, the inverse time
$a_{\rm t}=\dot{\Psi} (t_0)/2$ (\ref{2.10}) can be expressed as
\begin{equation} a_{\rm t}={H_0\over 2}\,{(1-9\Omega _\Lambda )\Phi
_0/c_0^2\over 1+2(1-3\Omega _\Lambda )\Phi
_0/c_0^2}\,.\label{4.2}\end{equation}
 Introducing in this
equation the values of $\Omega _\Lambda$ and $\Phi _0$, the
acceleration of the clocks $a_{\rm t}$ turns out to be
\begin{equation} a_t\simeq 0.8\, H_0 \simeq 1.8 \times
10^{-18}\mbox{ s}^{-1}\, .\label{4.3}\end{equation}

This is to be compared with the value suggested by Anderson {\it
et al} on the basis of their data $a_{\rm t}=(2.9\pm 0.3)\times
10^{-18}\mbox{ s}^{-1}$. Taking into account the simplicity of the
calculation and the approximations involved, the prediction of
this model can be considered to be acceptable. This is
encouraging.

\paragraph{4.1 An intuitive and phenomenological understanding of the
phenomenon.}

 Let a photon with frequency $h\nu _0$ travel across
the space at time $\tau _0$. Its energy is $h\nu _0$. If the time
runs from $\tau _0$ until $\tau$, when the gravitational potential
is $\Psi (\tau )$, it will pick up potential energy, its total
energy being then $h\nu (\tau )=h\nu _0 [1+\Psi (\tau)-\Psi (\tau
_0)]$. It will be seen as having a frequency $\nu (\tau )=\nu _0
[1+\Psi (\tau)-\Psi (\tau _0)]$. The derivative of the frequency
with respect to the time of the atomic clocks will be $2a_\tau
={\rm d}\Psi /{\rm d}\tau _{\rm clocks}
>0$. In other words, it must be expected that the frequency of the photons will increase
adiabatically, because of the expansion of the universe. This
means a small blueshift, just what was observed. Since $c^*$ is
constant, the wavelengths must decrease accordingly. That is the
Pioneer anomaly.

\section{SUMMARY AND CONCLUSIONS}
The model here presented suggests an explanation of the Pioneer
anomaly that is simple and based in standard physical ideas: {\it
the acceleration $a_{\rm P}$ is not related to any anomalous or
unmodelled motion of the spaceships. Instead, it is an effect of
the increasing background gravitational potential that pervades
the universe and produces an acceleration $2a_{\rm t}$ of the time
of the clocks $\tau _{\rm clocks}$ with respect to the coordinate
time $t$, {\it i. e.} $a_{\rm t}=\left.{\rm d}^2\tau _{\rm
clocks}/{\rm d} t^2\right|_{t_0}/2$,} $t_0$ being the present time
(remember that the time coordinate was so chosen as to go to the
Newtonian time in the Newtonian limit). This acceleration of the
clocks $2a_{\rm t}$ is also equal, in this model, to the time
derivative of the background gravitational potential $\Psi(t)$,
{\it i. e.} $a_{\rm t}=\left.{\rm d}\Psi /{\rm d}t\right|_{t_0}/2=
\left.{\rm d}\Psi /{\rm d}\tau _{\rm clocks}\right|_{t_0}/2$ (note
that ${\rm d}\tau _{\rm clock}/{\rm d}t =1$, at present time $t_0$
but not before or after). The anomaly would be thus an interesting
case of the dynamics of time$^{(20,\,21)}$. A further comment: as
it might be worth to point out, this model is similar to the
explanation by Mach of the origin of the inertia.

According to this model, the anomaly is a manifestation of the
expansion of the universe, which causes the increase of the
background potential $\Psi (t)$. This increase, in its turn,
accelerates the cosmological time and causes the acceleration
$2a_{\rm t}$. A simple estimation gives a good agreement with the
value proposed, on the basis of their observations, by Anderson
{\it et al}, the discoverers of the anomaly (see eq. (\ref{4.3}).)

The model complies with the principles of general relativity,
particularly because the proper speed of light ({\it i. e.}
$c^*={\rm d}\ell /{\rm d}\tau$, $\tau$ being the cosmological
proper time) is a universal constant. However, what is here called
the non-proper speed of light ({\it i. e.} $c(t)={\rm d}\ell /{\rm
d}t$, $t$ being the coordinate time) is not constant. This is
standard. If there is only matter, be it ordinary or dark, then
$c(t)<c^*$. On the other hand, it happens that $c(t)$ can
accelerate until being larger than the proper light speed $c^*$,
{\it if and only if there is cosmological constant (or any other
form of dark energy that implies repulsion)}. This is unusual and
might seem perplexing at first sight, but it does not imply any
contradiction as far as $c^*$ is still constant. It would be just
another unexpected effect of the dark energy, in addition to
accelerate the universe.

The Pioneer anomaly poses a most intriguing riddle for physics and
very complex and difficult problems for metrology. Taking this
into account, the main conclusion of this paper is that the ideas
here presented should be considered by the experts who know the
details of the motion of the spacecrafts and of the metrological
procedures involved in the observation.

\bigskip

{\bf ACKNOWLEDGEMENTS}

I am indebted to A. Tiemblo for information on the dynamics of
time, to J. L. Sebasti\'an for explanations on microwave detection
and to C. Aroca, A. I. G\'omez de Castro, E. L\'opez and J. L.
Rosales for discussions or encouragement. This research was
partially supported by Grant number BFM03-5453 of the Spanish
Ministry of Science and Technology.

\bigskip

{\bf REFERENCES}

{\small
\begin{enumerate}
\item  J. D. Anderson, Ph. A. Laing, E. L. Lau, A. S. Liu, M.
Martin Nieto  and S. G. Turyshev, {\it  Phys. Rev. Lett.} {\bf
81}, 2858 (1998).

\item J. D. Anderson, Ph. A. Laing, E. L. Lau, A. S. Liu, M.
Martin Nieto  and S. G. Turyshev, {\it Phys. Rev. D} {\bf 65},
082004 (2002).

\item J. D. Anderson, E. L. Lau, S. G. Turyshev, Ph. A. Laing and
M. Martin Nieto, {\it Mod. Phys. Lett. A} {\bf 17}, 875 (2002).

\item J. P. Mbelek and M. Michalski, {\it Int. J. Mod. Phys. D}
{\bf 13}, 865 (2004).

\item A. F. Ra\~nada, astro-phys/0202224, {\it Europhys. Lett.}
{\bf 61}, 174 (2003).

\item A. F. Ra\~nada, gr-qc/0211052, {\it Europhys. Lett.} {\bf
63}, 653 (2003)

\item A. F. Ra\~nada, {\it Int J. Mod. Phys. D} {\bf 12}, 1755
(2003).

\item A. F. Ra\~nada, gr-qc/0403013.

\item J. L. Rosales and J. L. S\'anchez-G\'omez, gr-qc/9810085.

\item J. L. Rosales, gr-qc/0401014.

\item G. F. R. Ellis and J. Ph. Uzan, gr-qc/0305099 v2.

\item C. W. Misner, K. S. Thorne and J. A. Wheeler, {\it
Gravitation} (Freeman, San Francisco, 1973).

\item M. V. Berry, {\it Principles of Cosmology and Gravitation}
 (Adam Hilger-Institute of Physics Publishing, Bristol, 1989).

\item  C. M. Will, {\it Theory and experiment in gravitational
physics} (Cambridge University Press, Cambridge, 1993), Rev. ed.

\item W. Rindler, {\it Relativity. Special, general and
cosmological} (Oxford University Press, Oxford, 2001).

\item L. D. Landau and E. M. Lifshitz, {\it The Classical Theory
of Fields}, 4th revised English edition (Pergamon Press, Oxford,
1975), chapter 10.

\item A. Einstein, {\it Annalen der Physik} {\bf 35}, 898 (1911);
reprinted in {\it The collected papers of Albert Einstein},
English translation, Vol 3 (Princeton University Press, 1993), p.
379.

\item A. Pais, {\it Subtle is the Lord. The science and the life
of Albert Einstein} (Oxford University Press, Oxford, 1984).

\item J. M. S\'anchez Ron, {\it El origen y desarrollo de la
relatividad} (Alianza Editorial, Madrid, 1985).

\item A. Tiemblo and R. Tresguerres, {\it Gen. Rel. Grav.} {\bf
34}, 31 (2002).

\item J. F. Barbero, A. Tiemblo and R. Tresguerres, {\it Phys.
Rev. D} {\bf 57}, 6104 (1998).

\end{enumerate}

\end{document}